\documentclass[aps,twocolumn,pra,superscriptaddress,amsmath,amssymb]{revtex4-1} %% for PRB
\usepackage{dcolumn}
\usepackage{bm}
\usepackage{amsmath}
\usepackage{txfonts}
\usepackage[T1]{fontenc}
\usepackage{xspace}
\usepackage{ulem}
\usepackage{multirow}
\usepackage{comment}
\ifx\pdfoutput\undefined
\usepackage[dvipdfmx]{graphicx}
\usepackage[dvipdfmx]{hyperref}
\usepackage[dvipdfmx]{color}
\else
\usepackage{graphicx}
\usepackage{hyperref}
\usepackage{color}
\fi
\usepackage{soul}
\soulregister\cite7
\soulregister\ref7
\soulregister\pageref7

\DeclareRobustCommand{\nchange}[2]{\ifmmode{{\textrm{\setul{}{1pt}\setstcolor{blue}\st{$\displaystyle#1$}}}}\else{{\setul{}{1pt}\setstcolor{blue}\st{#1}}}\fi\ \textcolor{blue}{#2}}

\begin{document}
\let\emph\textit

\title{
  Residual Entropy and Spin Fractionalizations in the Mixed-Spin Kitaev Model
}
\author{Akihisa Koga}
\affiliation{
  Department of Physics, Tokyo Institute of Technology,
  Meguro, Tokyo 152- 8551, Japan
}
\author{Joji Nasu}

\affiliation{
  Department of Physics, Yokohama National University,
  Hodogaya, Yokohama 240-8501, Japan
}

\date{\today}
\begin{abstract}
  We investigate ground-state and finite temperature properties of
  the mixed-spin $(s, S)$ Kitaev model.
  When one of spins is half-integer and the other is integer,
  we introduce two kinds of local symmetries,
  which results in a macroscopic degeneracy in each energy level.
  Applying the exact diagonalization to several clusters
  with $(s, S)=(1/2, 1)$,
  we confirm the presence of this large degeneracy in the ground states,
  in contrast to the conventional Kitaev models.
  By means of the thermal pure quantum state technique,
  we calculate the specific heat, entropy, and spin-spin correlations
  in the system.
  We find that in the mixed-spin Kitaev model with $(s, S)=(1/2, 1)$,
  at least, the double peak structure appears in the specific heat
  and the plateau in the entropy at intermediate temperatures,
  indicating the existence of the spin fractionalization.
  Deducing the entropy in the mixed-spin system with $s, S\le 2$ systematically,
  we clarify that the smaller spin-$s$ is responsible for
  the thermodynamic properties
  at higher temperatures.
\end{abstract}
\maketitle

%%%%%%%%%%%%%%%%%%%%%%%%%%%%%%%%%%%%%%%%%
%\section{Introduction}
%%%%%%%%%%%%%%%%%%%%%%%%%%%%%%%%%%%%%%%%%
Kitaev model~\cite{KITAEV20062} and its related models
have attracted much interest in condensed matter physics
since the possibility of the direction-dependent Ising interactions
has been proposed in the realistic materials~\cite{Jackeli}.
Among them, low temperature properties
in the candidate materials such as
$A_2{\rm IrO_3}$ ($A={\rm Na,  K}$)~\cite{PhysRevB.82.064412,PhysRevLett.108.127203,PhysRevLett.108.127204,PhysRevLett.109.266406,modic2014realization,PhysRevLett.114.077202,Kitagawa2018nature}
and $\alpha$-${\rm RuCl_3}$~\cite{PhysRevB.90.041112,Kubota,PhysRevB.91.144420,PhysRevB.91.180401,Kasahara}
have been examined extensively.
To clarify the experimental results,
the roles of the Heisenberg interactions~\cite{Chaloupka,Jiang,Singh},
off-diagonal interactions~\cite{Katukuri,Suzuki},
interlayer coupling~\cite{Tomishige,Seifert,Tomishige2}, and
the spin-orbit couplings~\cite{Nakauchi} have been theoretically investigated for both ground state and finite temperature properties.
One of the important issues characteristic of the Kitaev models is the fractionalization
of the spin degree of freedom.
In the Kitaev model with $S=1/2$ spins, the spins are exactly shown to be fractionalized
into itinerant Majorana fermions and localized fluxes, which
manifest themselves in the ground state and thermodynamic properties~\cite{Nasu1,Nasu2}.
It has been observed as the half-quantized thermal quantum Hall effects,
which is a clear evidence of the Majorana quasiparticles fractionalized from quantum spins~\cite{Kasahara}.
Recently, the Kitaev model with larger spins has theoretically
been examined~\cite{Baskaran,SuzukiYamaji,S1Koga,Oitmaa,Kee}.
In the spin-$S$ Kitaev model,
the specific heat exhibits double peak structure,
and plateau appears in the temperature dependence of the entropy~\cite{S1Koga}.
This suggests the existence of the fractionalization
even in this generalized Kitaev model.
However, it is still hard to explain
how the spin degree of freedom is divided in the generalized Kitaev models
beyond the exactly solvable $S=1/2$ case~\cite{KITAEV20062,Nasu1,Nasu2}.

The key to understand the ``fractionalization'' in the spin-$S$ Kitaev model should be
the multiple entropy release phenomenon.
The half of spin entropy $\sim \frac{1}{2}\ln (2S+1)$
in higher temperatures emerges with a broad peak in the specific heat.
Then, a question arises how the plateau structure appears in the entropy
in the Kitaev model composed of multiple kinds of spins
(the mixed-spin Kitaev model).
In other words, how is the many-body state realized in the system,
with decreasing temperatures?
The extension to the mixed-spin models should be a potential
to exhibit an intriguing nature of the ground states.
In fact, the mixed-spin quantum Heisenberg model has been
examined~\cite{Mitsuru,Pati,Fukui,Tonegawa,KogaMix,KogaMix2,Kolezhuk,Takushima},
and the topological nature of spins and lattice plays an important role
in stabilizing the non-magnetic ground states.
Moreover, mixed-spin Kitaev model can be realized
by replacing transition metal ions to other ions
in the Kitaev candidate materials.
Therefore, it is desired to study this model
to discuss the nature of the spin fractionalization in the Kitaev system.

%%%%%%%%%%%%%%%%%%%%%%%%%%%%%%%%%%%%%%%%%%
\begin{figure}[htb]
  \centering
  \includegraphics[width=8cm]{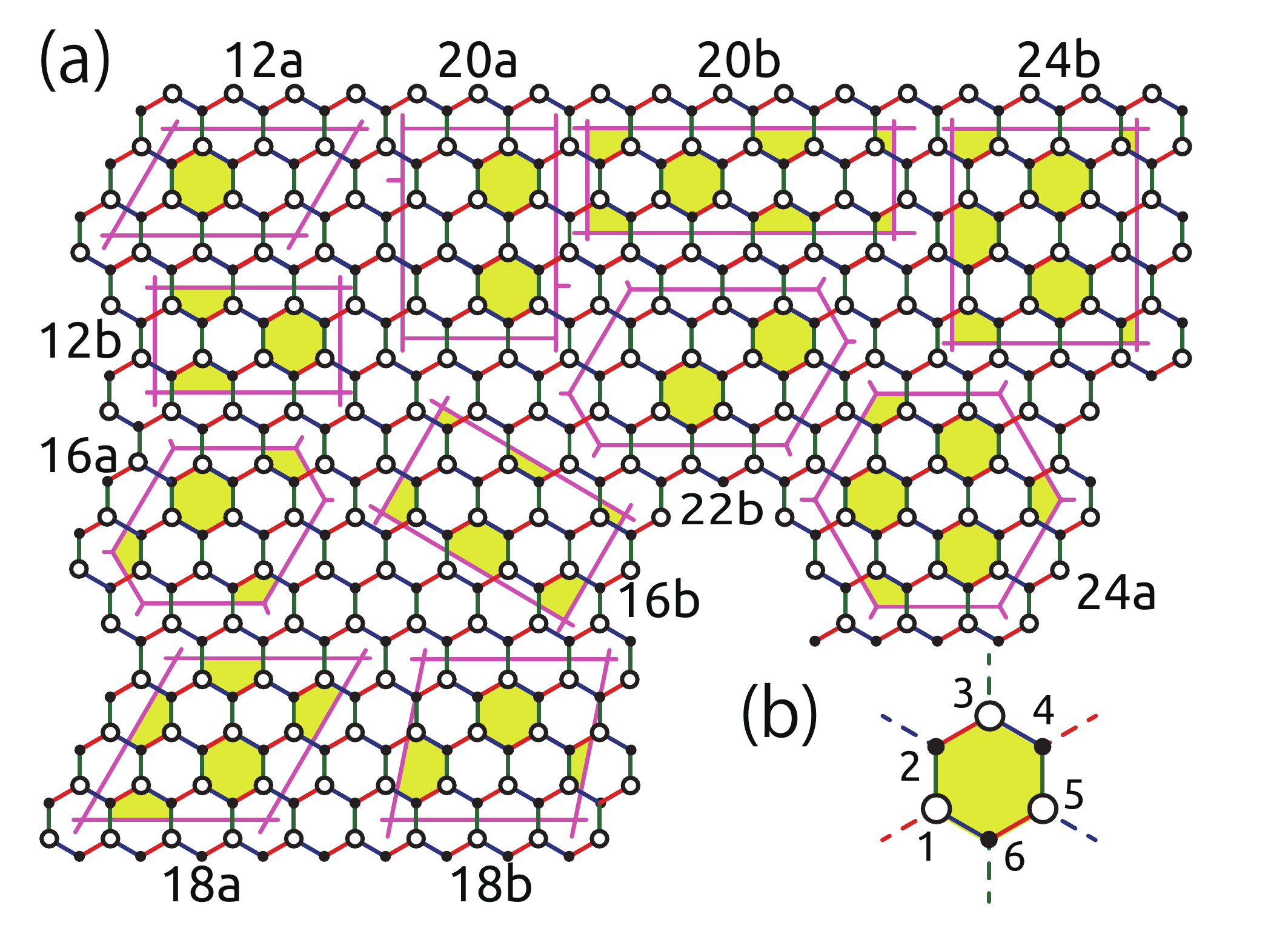}
  \caption{
    (a) Mixed-spin Kitaev model on a honeycomb lattice.
    Solid (open) circles represent spin $s$ ($S$).
    Red, blue, and green lines denote $x$, $y$, and $z$ bonds
    between nearest neighbor sites, respectively.
    (b) Plaquette with sites marked $1-6$ is shown
    for the corresponding operator
    $W_p$ defined in Eq.~(\ref{eq:Wp}).
  }
  \label{fig:model}
\end{figure}
%%%%%%%%%%%%%%%%%%%%%%%%%%%%%%%%%%%%%%%%%%

In this manuscript, we investigate the mixed-spin Kitaev model,
where two distinct spins $(s, S)\;[s<S]$ are periodically arranged
on the honeycomb lattice (see Fig.~\ref{fig:model}).
First, we show the existence of the $Z_2$ symmetry in each plaquette in the system.
In addition, by considering another local symmetry,
we show that the macroscopic degeneracy exists in each energy level
when one of the spins is half-integer and the other integer.
The exact diagonalization (ED) in the system
with $(s,S)=(\frac{1}{2},1)$ reveals that
the ground state has a macroscopic degeneracy,
which is consistent with the presence of the two kinds of local symmetries.
Using thermal pure quantum (TPQ) state methods~\cite{TPQ1,TPQ2},
we find that, at least, the double peak structure appears
in the specific heat and
the plateau appears at intermediate temperatures in the entropy,
which are similar to those in the spin-$S$ Kitaev models~\cite{S1Koga}.
From systematic calculations for the mixed-spin systems with $s, S\le 2$,
we clarify that the smaller spin-$s$ is responsible
for the high-temperature properties.
The deconfinement picture to explain the ``spin fractionalization''
in the Kitaev model is addressed.

%The paper is organized as follows.
%In Sec.~\ref{sec:model},
%we introduce the generalized Kitaev model and briefly summarize
%our numerical techniques.
%In Sec.~\ref{sec:results},
%we study ground-state and finite temperature properties in the model.
%A summary is given in the final section.

%%%%%%%%%%%%%%%%%%%%%%%%%%%%%%%%%%%%%
%\section{Model and Method}\label{sec:model}
%%%%%%%%%%%%%%%%%%%%%%%%%%%%%%%%%%%%%
We consider the Kitaev model on a honeycomb lattice,
which is given by the following Hamiltonian as
%%%%%%%%%%%%%%%%%%%%%%%%%%%%
\begin{eqnarray}
  {\cal H} &=&
  -J\sum_{\langle i,j \rangle_x}s_i^x S_j^x
  -J\sum_{\langle i,j \rangle_y}s_i^y S_j^y
  -J\sum_{\langle i,j \rangle_z}s_i^z S_j^z,\label{eq:H}
\end{eqnarray}
%%%%%%%%%%%%%%%%%%%%%%%%%%%%%
where $s_i^\alpha(S_i^\alpha)$ is the $\alpha(=x,y,z)$ component
of a spin-$s(S)$ operator at the $i$th site.
$J$ is the exchange constant between the nearest neighbor spin pairs $\langle i,j \rangle_\gamma$.
The model is schematically shown in Fig.~\ref{fig:model}(a).
We consider here the following
local Hermite operator defined on each plaquette $p$ as,
\begin{eqnarray}
  W_p &=& \exp\Big[i\pi \left(S_1^x+s_2^y+S_3^z+s_4^x+S_5^y+s_6^z\right)-i\pi\eta \Big],\label{eq:Wp}
\end{eqnarray}
where $\eta=[3(s+S)]$ is a phase factor.
%This operator commutes with the Hamiltonian Eq.~(\ref{eq:H}), and
%$W_p^2=1$ is satisfied.
By using the following relation for the spin operators,
%\begin{equation}
%  e^{i\pi S^\alpha}S^\beta e^{-i\pi S^\alpha}=\left\{
%  \begin{array}{ll}
%    -S^\beta & (\alpha\neq\beta)\\
%    S^\beta  & (\alpha=\beta)
%  \end{array}
%  \right.,\label{eq:exp}
%\end{equation}
%\begin{eqnarray}
$  e^{i\pi S^\alpha}S^\beta e^{-i\pi S^\alpha}=(2\delta_{\alpha\beta}-1)S^\beta,$
%  \label{eq:exp}
%\end{eqnarray}
we find $[{\cal H}, W_p]=0$ for each plaquette and $W_p^2=1$.
Therefore, the mixed-spin Kitaev system has a $Z_2$ local symmetry.
%Namely, Eq.~(\ref{eq:Wp}) is a natural extension of the local operator
%in the generalized spin-$S$ Kitaev model~\cite{Baskaran}.

It is known that this local $Z_2$ symmetry is important to
understand ground state properties in the Kitaev model.
%In the exactly solvable $S=1/2$ Kitaev model,
%the ground state belongs to the manifold with $w_p=+1$ for all plaquettes,
%where $w_p(=\pm 1)$ is an eigenvalue of $W_p$,
%and a massless fermionic dispersion appears in the elementary excitations.
%In the $S=1$ Kitaev model, it has been numerically confirmed that $w_p=+1$
%is satisfied for all plaquettes in the ground state.
%Therefore, it is instructive to clarify the distribution of the quantities
%in the ground state of the mixed-spin Kitaev model.
We wish to note that the local operator $W_p$ on a plaquette $p$
commutes with those on all other plaquettes
in the spin-$S$ Kitaev models,
while this commutation relation is not always satisfied
in the present mixed-spin Kitaev model.
In fact, %by using Eqs.~(\ref{eq:Wp}) and (\ref{eq:exp}),
we obtain $[W_p, W_q]\propto [e^{i\pi(s_i^x+S_j^y)},e^{i\pi(s_i^y+S_j^x)}]\propto
\sin[\pi(s_i^y-S_j^x)]$
when the plaquettes $p$ and $q$ share the same $z$ bond $\langle ij\rangle_z$.
This means that the local operator does not commute with the adjacent ones
in the mixed-spin Kitaev model with one of spins being half-integer and
the other integer.
Instead, we introduce another local symmetry specific in this case.
When either $s$ or $S$ is half-integer and the other is integer,
the Hilbert space is divided into subspaces specified by
the set of the eigenvalues $w_p(=\pm 1)$ of
the $N_p(\le N/6)$ local operators $W_p$
defined on the plaquettes $p\in {\cal P}$,
where ${\cal P}$ is a set of the plaquettes whose corners
are not shared with each other.
Now, we assume the presence of the local operator $R_p$ on a plaquette $p(\in {\cal P})$
so as to satisfy the conditions,
$R_p^2=1$ and the following commutation relations
$[{\cal H}, R_p]=0$, $[W_p, R_q]=0\; (p\neq q)$, and $\{W_p, R_p\}=0$.
In the case that half-integer and integer spins are mixed,
such an operator can be introduced so that
the spins located on its corners
are inverted as
${\bf S}_{2i-1}\rightarrow -{\bf S}_{2i-1},
{\bf s}_{2i}\rightarrow -{\bf s}_{2i}\;(i=1,2,3)$
and the signs of the six exchange constants are changed on the bonds connecting with a corner site belonging to the plaquette,
shown as the dashed lines in Fig.~\ref{fig:model}(b).
When a wavefunction for the energy level $E$ is given by the set of $\{w_p\}$ as
$|\psi\rangle=|\psi;\{w_1, w_2,\cdots,w_p,\cdots\}\rangle$,
we obtain ${\cal H}|\psi'\rangle=E|\psi'\rangle$ with the wave function
$|\psi'\rangle=R_p|\psi\rangle=|\psi;\{w_1, w_2,\cdots,-w_p,\cdots\}\rangle$.
Since the operators $R_p$ for arbitrary plaquettes in ${\cal P}$ generates
degenerate states, the presence of $R_p$ results in,
at least, $2^{N/6}$-fold degenerate ground states.

This qualitative difference in the spin magnitudes
$s$ and $S$ can be confirmed in the small clusters.
By using the ED method, we obtain ground state properties
in the twelve-site systems, as shown in Table~\ref{tbl2}.
We clearly find that, as for the ground-state degeneracy, the mixed-spin systems
can be divided into three groups.
When both spins $s$ and $S$ are integer,
the ground state is always singlet.
In the half-integer case,
the four-fold degenerate ground state is realized in the $N=12a$ system,
while
the singlet ground state is realized in the $N=12b$ system.
This feature is essentially the same as ground state properties
in the $S=1/2$ Kitaev model,
where the ground-state degeneracy depends on the topology in the boundary condition.
By contrast,
the eight-fold degenerate state is realized
in the system with one of spins being half-integer and
the other integer,
which suggests the macroscopic degeneracy in the thermodynamic limit.

%%%%%%%%%%%%%%%%%%%%%%%%%%%%%%%%%%%%%%%%%
\begin{center}
\begin{table}
\caption{
  Ground state energy $E_g$ and its degeneracy $N_d$
  in the mixed-spin $(s, S)$ Kitaev models with the twelve-site clusters.
}

\begin{tabular}{cc|cc|cc}
  \hline \hline
  \multirow{2}{*}{$s$} & \multirow{2}{*}{$S$} &\multicolumn{2}{c|}{$N=12a$}&\multicolumn{2}{c}{$N=12b$}\\
  && $E_g/JN$ & $N_d$ & $E_g/JN$ & $N_d$\\
  \hline
  1/2  & 1/2 & -0.20417 & 4 & -0.21102 & 1  \\
  1/2  &  1  & -0.33533 & 8 & -0.34235 & 8  \\
  1/2  & 3/2 & -0.47208 & 4 & -0.47389 & 1  \\
  1/2  &  2  & -0.60214 & 8 & -0.60260 & 8  \\
  1   &  1   & -0.66487 & 1 & -0.67421 & 1  \\
  1   & 3/2 &  -0.92855 & 8 & -0.93437 & 8  \\
  1   &  2  &  -1.19271 & 1 & -1.19567 & 1  \\
  3/2 & 3/2 &  -1.37169 & 4 & -1.38840 & 1  \\
  3/2 &  2  &  -1.76901 & 8 & -1.77691 & 8  \\
  2   & 2   &  -2.33449 & 1 & -2.35306 & 1  \\
  \hline \hline
\end{tabular}

\label{tbl2}
\end{table}
\end{center}
%%%%%%%%%%%%%%%%%%%%%%%%%%%%%%%%%%%%%%%%%

%%%%%%%%%%%%%%%%%%%%%%%%%%%%%%%%%%%%%
%\section{Results} \label{sec:results}
%%%%%%%%%%%%%%%%%%%%%%%%%%%%%%%%%%%%%
To confirm this,
we focus on the mixed-spin system with $(s, S)=(1/2, 1)$.
By using the ED method,
we obtain the ground-state energies for several clusters up to 24 sites
[see Fig.~\ref{fig:model}(a)].
The obtained results are shown in Table~\ref{tbl}.
It is clarified that a finite size effect slightly appears
in the ground state energy, and
its value is deduced as $E_g/JN=-0.335$.
We also find that the ground state is $N_{\cal S}(=N_d/2^{N_p})$-fold degenerate
in each subspace
and its energy is identical in all subspaces ${\cal S}[\{w_p\}]$
except for the $N=18a$ system~\cite{N18a}.
The large ground-state degeneracy $N_d\ge 2^{N/6}$ is consistent with
the above conclusion.
We also find that the first excitation energy $\Delta$ is much smaller than
the exchange constant $J$,
as shown in Table~\ref{tbl}.
These imply the existence of multiple low-energy states in the system.

%%%%%%%%%%%%%%%%%%%%%%%%%%%%%%%%%%%%%%%%%
\begin{center}
\begin{table}
\caption{
  Ground state profile for several clusters in the Kitaev model
  with $(s, S)=(1/2, 1)$. $N_p$ is the number of plaquettes,
  where the local operator $W_p$ is diagonal in the basis set.
  $N_d$ is the degeneracy in the ground state.
}
\begin{tabular}{ccccc|ccccc}
  \hline \hline
$N$ & $N_p$ & $E_g/JN$ & $\Delta/J$ & $N_d$ & $N$ & $N_p$ & $E_g/JN$ & $\Delta/J$ & $N_d$  \\
  \hline
  12a & 1 & -0.33981 & 0.0071 & 8 & 20a & 2 & -0.33550 & 0.0013 &20 \\
  12b & 2 & -0.34235 & 0.0024 & 8 & 20b & 3 & -0.34210 & 0.0041 &32 \\
  16a & 2 & -0.33543 & 0.0002 &20 & 22  & 2 & -0.33531 & 0.0016 &20 \\
  16b & 2 & -0.33895 & 0.0019 &16 & 24a & 4 & -0.33525 & 0.0031 &64 \\
  18a & 3 & -0.33533 & 0.0018 &8  & 24b & 4 & -0.33511 & 0.0010 &64 \\
  18b & 2 & -0.33537 & 0.0015 &40 \\
  \hline \hline
\end{tabular}
\label{tbl}
\end{table}
\end{center}
%%%%%%%%%%%%%%%%%%%%%%%%%%%%%%%%%%%%%%%%%

Next, we consider thermodynamic properties in the Kitaev model.
It is known that there exist two energy scales in the $S=1/2$ Kitaev model~\cite{KITAEV20062},
which clearly appear as
double peak structure in the specific heat
and a plateau in the entropy~\cite{Nasu1,Nasu2}.
Similar behavior has been reported in the spin-$S$ Kitaev model~\cite{S1Koga}.
These suggest the existence of the fractionalization
in the generalized spin-$S$ Kitaev model.
An important point is that the degrees of freedom for the high energy part
depend on the magnitude of spins $\sim(2S+1)^{N/2}$.
On the other hand, in the mixed-spin case, it is unclear which spin is responsible
for the high-temperature properties.

%To clarify the nature of the fractionalization in the Kitaev model,
%we examine finite-temperature properties in the mixed-spin model.
Here, we calculate thermodynamic quantities for twelve-site clusters,
by diagonalizing the corresponding Hamiltonian.
Furthermore, we apply the TPQ state method~\cite{TPQ1,TPQ2} to larger clusters.
In this calculation, the thermodynamic quantities are deduced
by the statistical average of the results obtained from, at least,
25 independent TPQ states.
Here, we calculate specific heat $C(T)=dE(T)/dT$,
entropy $S(T)=S_\infty -\int_T^\infty C(T')/T' dT'$, and
the nearest-neighbor spin-spin correlation
$C_S(T)=\langle s_i^\alpha S_j^\alpha \rangle_\alpha = -2E(T)/(3J)$,
%which are defined as,
%\begin{eqnarray}
%  C(T) &=& \frac{dE(T)}{dT},\\
%  S(T) &=& S_\infty -\int_T^\infty \frac{C(T')}{T'} dT',\\
%  C_s(T) &=& \langle s_i^\alpha S_j^\alpha \rangle_\alpha = -\frac{2E(T)}{3J},
%\end{eqnarray}
where $S_\infty= \frac{1}{2} \ln (2s+1)(2S+1)$ and $E(T)$ is the internal energy per site.
%%%%%%%%%%%%%%%%%%%%%%%%%%%%%%%%%%%%%%%%%%
\begin{figure}[htb]
  \centering
  \includegraphics[width=8cm]{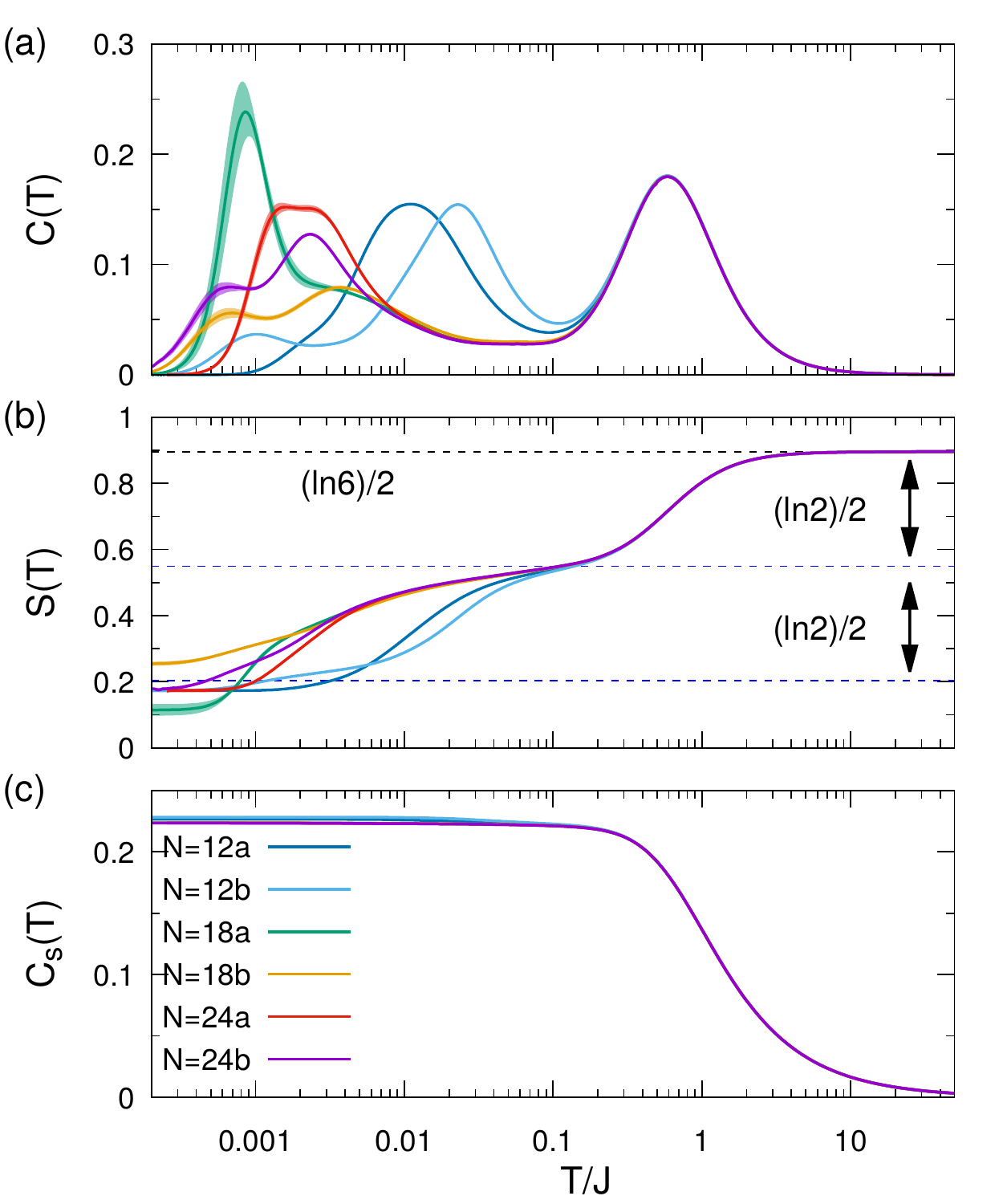}
  \caption{
    (a) Specific heat, (b) entropy, and (c) spin-spin correlation
    as a function of temperatures.
    Shaded areas stand for the standard deviation of the results
    obtained from the TPQ states.
  }
  \label{fig:tpq}
\end{figure}
%%%%%%%%%%%%%%%%%%%%%%%%%%%%%%%%%%%%%%%%%%

The results for the mixed-spin systems with $(s, S)=(1/2, 1)$
are shown in Fig.~\ref{fig:tpq}.
We clearly find the multiple-peak structure in the specific heat.
Note that
finite size effects appear only at low temperatures.
Therefore, our TPQ results for the 24 sites appropriately capture
the high temperature properties ($T\gtrsim 0.01J$) in the thermodynamic limit.
Then, we find the broad peak around $T_H\sim 0.6J$,
which is clearly separated by the structure at low temperatures ($T<0.01J$).
%To clarify the nature of the thermodynamic properties at high temperatures,
Now, we focus on the corresponding entropy,
which is shown in Fig.~\ref{fig:tpq}(b).
This indicates that, decreasing temperature,
the entropy monotonically decreases and the plateau structure is found
around $T/J\sim 0.1$.
The released entropy is $\sim\frac{1}{2}\ln 2$,
which is related to the smaller spin $(s=1/2)$.
Therefore, %although the system is composed of two kinds of spins $(s, S)$,
multiple temperature scales do not appear at high temperatures
although the system is composed of two kinds of spins $(s, S)$.
However, it does not imply that only smaller spins are frozen
and larger spins remain paramagnetic at the temperature
since the spin-spin correlations develop around $T\sim T_H$ and
a quantum many-body spin state %without the magnetic moments
is formed, as shown in Fig.~\ref{fig:tpq}(c).
We have also confirmed that local magnetic moments do not appear even
in the wavefunction constructed by the superposition of the ground states
with different configurations of $\{w_p\}$.

By contrast,
the value $\frac{1}{2}\ln 2$ reminds us of the high-temperature feature for
itinerant Majorana fermions in spin-1/2 Kitaev model~\cite{Nasu1,Nasu2}.
Then, one expects that, in the mixed-spin $(s, S)$ Kitaev model,
higher temperature properties are described by the smaller spin-$s$ Kitaev model,
where degrees of freedom $\sim (2s+1)^{1/2}$ are frozen at each site~\cite{S1Koga}.
In the case, a peak structure appears in the specific heat and
the plateau structure at $\sim S_\infty - \ln (2s+1)/2$ in the entropy.
These interesting properties at higher temperatures will be examined systematically.
Further decrease of temperatures decreases the entropy and
finally $S \sim S_\infty-\ln 2$ at lower temperatures, as shown in Fig.~\ref{fig:tpq}(b).
This may suggest that thermodynamic properties in this mixed-spin Kitaev model
with $(s, S)=(1/2, 1)$ are governed by two kinds of fractional quasiparticles originating
from the smaller $s=1/2$ spin by analogy with the spin fractionalization
in the spin-1/2 Kitaev model.
In the case, the existence of the remaining entropy $S \sim S_\infty-\ln 2$
should be consistent with macroscopic degeneracy in the ground state
as discussed before.
However, our TPQ data have a large system size dependence at low temperatures,
and conclusive results could not be obtained.
Therefore, a systematic analysis is desired to clarify the nature of
low temperature properties.

\begin{figure}[htb]
  \centering
  \includegraphics[width=8cm]{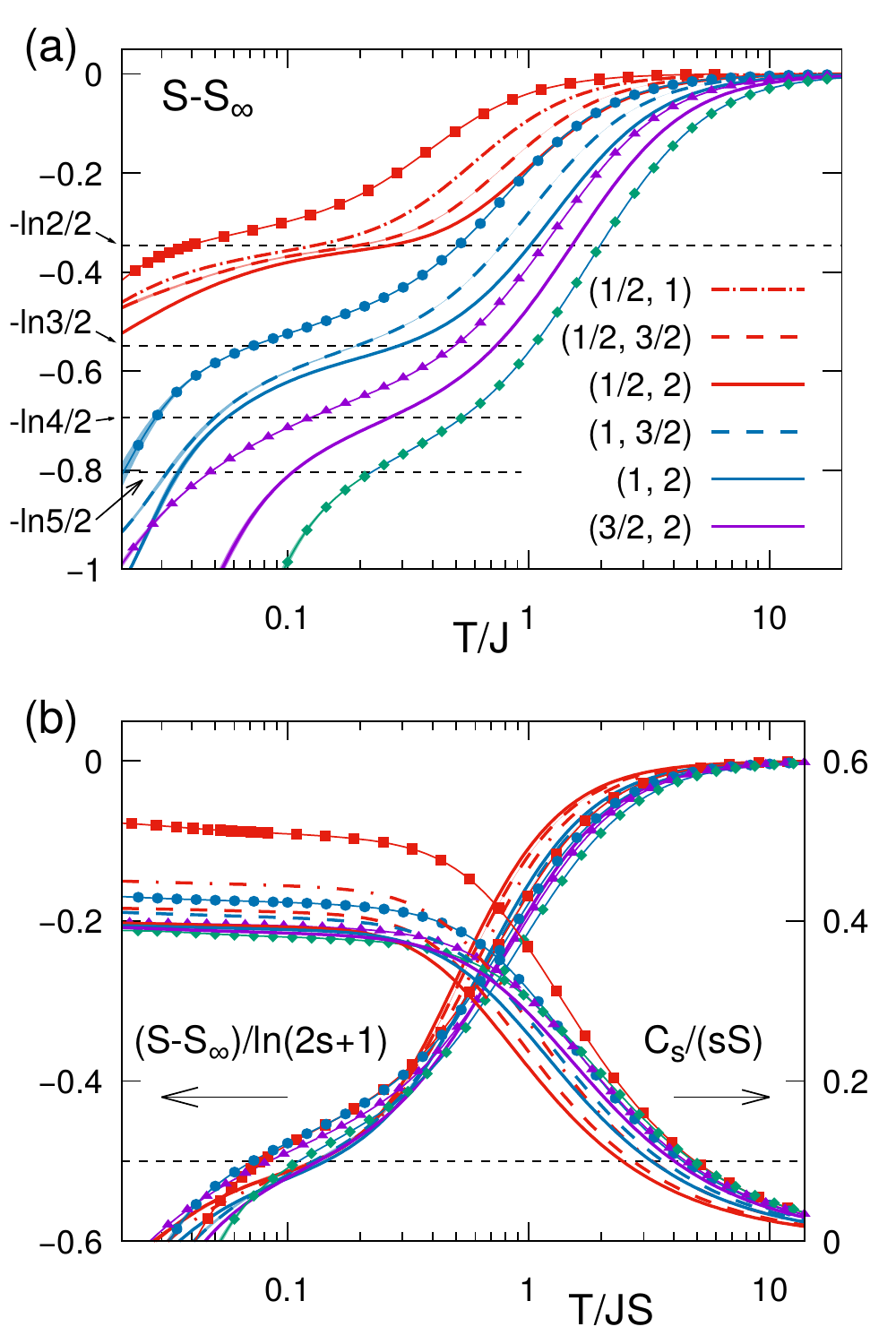}
  \caption{
    (a) $S-S_\infty$
    in the generalized $(s, S)$ Kitaev model
    at higher temperatures.
    Squares with lines represent data for the $S=1/2$ Kitaev model
    obtained from the Monte Carlo simulations~\cite{Nasu1,Nasu2}.
    Circles, triangles, and diamonds with lines represent
    the TPQ data for $S=1, 3/2$, and $2$ cases~\cite{S1Koga}.
    (b) $(S-S_\infty)/\ln (2s+1)$ and $C_s/(sS)$ as a function of $T/JS$.
  }
  \label{fig:tpqS}
\end{figure}

To clarify the role of the smaller spins in the mixed-spin Kitaev models,
we calculate the entropy in the systems with $s, S\le 2$ and $N=12a$
by means of the TPQ state methods.
The results are shown in Fig.~\ref{fig:tpqS}.
%%%%%%%%%%%%%%%%%%%%%%%%%%%%%%%%%%%%%%%%%%
%%%%%%%%%%%%%%%%%%%%%%%%%%%%%%%%%%%%%%%%%%
The plateau structure is clearly observed
in the curve of the entropy in the mixed-spin Kitaev models.
In addition, we find that the plateau is located around
$S=S_\infty-\frac{1}{2}\ln (2s+1)$, as expected above.
Therefore, we can say that, decreasing temperatures,
the half of the degree of freedom in the smaller spin-$s$ are released.

This may be explained by the deconfined-spin picture in the Kitaev model.
%which is exactly shown in the $S=1/2$ case.
In the picture, each spin $S$ is divided into two kinds of quasiparticles
with distinct energy scales: $2S$ $L$-quasiparticles and $2S$ $H$-quasiparticles,
which are dominant at lower and higher temperatures, respectively.
In the exactly solvable $S=1/2$ Kitaev model,
$H$- ($L$-)quasiparticles are identical to
itinerant Majorana fermions (localized fluxes).
In addition, this should explain the double peak structure
in the specific heat of the spin-$S$ Kitaev model,
each of which corresponds to the half entropy release~\cite{S1Koga}.
In our mixed-spin $(s,S)$ system,
the entropy release at higher temperatures can be interpreted as follows:
$2s$ fractional $H$-quasiparticles are present
with the energy scale of $\sim J$.
On the other hand, remaining $H$-quasiparticles originating from
larger spin-$S$ posses the energy that is much smaller than $J$
due to the absense of the two-dimensional network.
Therefore, only $2s$ $H$-quasiparticles form the many-body state
at high temperatures,
resulting in the plateau structure in the entropy.

Interestingly, the temperature $T^*$ characteristic of the plateau
in the entropy,
which may be defined such that $S(T^*)=S_\infty-\frac{1}{2}\ln (2s+1)$,
depends on the magnitude of the larger spin.
In fact, we find that
$T^*$ should be scaled by the larger spin $T^*\sim JS$,
which is shown in Fig.~\ref{fig:tpqS}(b).
This is in contrast to the conventional temperature scale
$T^{**}\sim J\sqrt{s(s+1)S(S+1)}$,
which is derived from the high-temperature expansion.
This discrepancy is common to the spin-$S$ Kitaev model~\cite{S1Koga},
implying that quantum fluctuations are essential even in this temperature range
in the mixed-spin Kitaev models.
%In the $S=1/2$ Kitaev model, thermal fractionalization occurs around $T^{**}$
%due to the quantum many-body effect, and therefore,
%the deviation from high-temperature expansion observed in the mixed-spin models
%may support the presence of the spin fractionalization.
As for the spin-spin correlation, decreasing temperatures,
it develops around $T/JS\sim 1$ and is almost saturated around $T^*$,
as shown in Fig.~\ref{fig:tpqS}(b).
This means that the many-body spin state is indeed realized at the temperature.
We also find that at low temperatures, the normalized spin-spin correlation $C_s/(sS)\sim 0.4$
is less than unity when $s$ and $S$ are large.
This suggests that the quantum spin liquid state is, in general, realized
in the generalized mixed-spin Kitaev model,
which is consistent with the presence of
magnetic fluctuations even in the classical limit~\cite{SuzukiYamaji}.

%%%%%%%%%%%%%%%%%%%%%%%%%%%%%%%%%%%%%%%%%%
%\begin{figure}[htb]
%  \centering
%  \includegraphics[width=8cm]{pic.pdf}
%  \caption{
%    Schematic pictures for the generalized Kitaev models.
%  }
%  \label{fig:pic}
%\end{figure}
%%%%%%%%%%%%%%%%%%%%%%%%%%%%%%%%%%%%%%%%%%

%%%%%%%%%%%%%%%%%%%%%%%%%%%%%%%%%%%%%
%\section{Summary}
%%%%%%%%%%%%%%%%%%%%%%%%%%%%%%%%%%%%%
In summary, we have studied the mixed-spin Kitaev model.
First, we have clarified the existence of the local $Z_2$ symmetry
at each plaquette.
We could introduce an operator $R_p$ on the plaquette $p$ so as
to (anti)commute with the Hamiltonian ($W_p$),
which leads to
the macroscopic degeneracy for each energy level
in the mixed-spin system with one of spins being half-integer and
the other integer.
%Examining the mixed-spin model with $(s, S)=(1/2, 1)$ numerically,
%we have confirmed that the system has
%the large degeneracy in the ground states.
Using the TPQ state methods for several clusters,
we have found the double peak structure in the specific heat and
plateau in the entropy, which suggests
the existence of the fractionalization in the mixed-spin system.
Deducing the entropy in the mixed-spin system with $s, S\le 2$ systematically,
we have clarified that the smaller spin plays a crucial role
in the thermodynamic properties at higher temperatures.
We expect that the present mixed-spin Kitaev systems are realizable in the real materials
by substituting the magnetic ions in the Kitaev candidate materials to other magnetic ions with larger spins,
and therefore, the present work should stimulate material researches
for mixed-spin Kitaev systems.

\begin{comment}
  The local $Z_2$ symmetry in the honeycomb lattice
exists even in the random-spin and random-bond Kitaev model,
%%%%%%%%%%%%%%%%%%%%%%%%%%%%
$ {\cal H} =\sum_{\langle i,j\rangle_x}J_{ij}S_i^x S_j^x+\sum_{\langle i,j\rangle_y}J_{ij}S_i^y S_j^y+
\sum_{\langle i,j\rangle_z}J_{ij}S_i^z S_j^z$,
%%%%%%%%%%%%%%%%%%%%%%%%%%%%%
where $\langle i,j\rangle_\alpha$ is the nearest neighbor pairs on the $\alpha$ bond,
as shown in Fig.~\ref{fig:model}(a).
In fact, the Hamiltonian commutes with the following local operator
$  W_p = \exp\Big[i\pi \left(S_1^x+S_2^y+S_3^z+S_4^x+S_5^y+S_6^z\right)-i\pi\eta\Big],$
where $\eta(=\sum_{i=1}^6 S_i)$ is a phase factor.
Therefore, it is naively expected that a non-magnetic ground state is realized
even in the random-spin and random-bond Kitaev model.
It is also interesting to clarify how robust thermodynamic properties characteristic of
the Kitaev model are against the introduction of magnetic impurities,
which is left for future work.
\end{comment}

\begin{acknowledgments}
  %  We thank ***.
  Parts of the numerical calculations were performed
  in the supercomputing systems in ISSP, the University of Tokyo.
  This work was supported by Grant-in-Aid for Scientific Research from
  JSPS, KAKENHI Grant Nos. JP18K04678, JP17K05536 (A.K.),
  JP16K17747, JP16H02206, JP18H04223 (J.N.).
\end{acknowledgments}

\bibliography{./refs}

\begin{thebibliography}{41}
\bibitem [{N18()}]{N18a}%
  \BibitemOpen
  \href@noop {} {}\bibinfo {note} {In the $N=18a$ system, the ground state
  belongs to two subspaces with all eigenvalues of $W_p$ being $\pm 1$. %
  Eigenstates with inharmonious distributions of $w_p$ are not the ground
  states. This distribution of $w_p$ is quite different from those in the other
  clusters. We expect that the local operators $R_p$ cannot be defined %in the
  $N=18a$ system because of the periodic boundary condition.}\BibitemShut
  {Stop}%
\end{thebibliography}

\end{document}